# MEASURING CREATIVITY OF WIKIPEDIA EDITORS

Pentti Launonen[a], KC Kern[b], Sanna Tiilikainen[c]

a,c Aalto University, School of Business, Runeberginkatu 14-16, Helsinki 00100, Finland
b Massachusetts Institute of Technology, 77 Massachusetts Avenue, Cambridge, MA 02139-4307, USA
e-mail: pentti.launonen@aalto.fi

**ABSTRACT**

Our paper explores contribution patterns of creativity and collaboration of Wikipedia editors as manifestations of social dynamics between the editors. We find support for existence of four socially constructed personas among the editors and difference in distribution of personas in articles of different qualities.

## 1. INTRODUCTION

Our paper explores how the patterns of contributions and collaboration of voluntary co-creation on virtual platforms are spontaneously organized as socially conditioned responses. We adopt the definition of creativity as "an action resulting in something novel and valuable" (Amabile, 1996), in this case Wikipedia articles. We focus on patterns of the highest quality ranking, called "Featured Article", and non-rated articles, called "Non-Assessed Articles".

The previous research in relation to creativity, teamwork and social roles in Wikipedia have concentrated on distinct editor types (e.g. Welser et al., 2008, Iba et al., 2009), personal editing behaviors (Liu and Ram, 2011) or group dynamics (Nemoto et al., 2011, Kittur, 2008, Kidane and Gloor, 2007). To our best knowledge, the studies have not combined the editing behaviors with the editing personalities to form a picture of the interplay between individual and social dynamics related to the creative editing behaviors in Wikipedia, the aim of this research.

## 2. DATA AND METHODS

Our research consisted of two qualitative rounds and one quantitative analysis: first, hypothetically identifying personas in terms of editing behavior; second, testing of existence of such personas in a larger scale; and third, statistical analysis of the tests.

For the first round of the qualitative research, we built a tool for collecting data from the editors of any given English Wikipedia article. The tool visualizes the contribution of top editors of the given Wikipedia article and plots the number of edits by annual quarters. Further, the tool visualizes the changes in the contributions of the editors from one quarter to the next, i.e. the first order derivative of edit volumes, and thus enables identifying aligned editing behaviors among editors. Furthermore, the tool calculates correlations between editing activities of each editor for the article, for the purpose of identifying alignment or disalignment of the editors. In this manner, we identified four distinct personas.

For the second round, we first randomly chose 115 non-quality-rated articles (NA) from Wikipedia, with the functionality of Wikipedia to read a random article; after omitting stub articles or articles with fewer than 50 page edits or 10 authors, this yielded 20 articles for the analysis. Second, we chose randomly 20 Featured Articles (FA) from Wikipedia, each representing one individual category in Wikipedia. We utilized the tool developed for the first round to test whether the detected personas are generalizable in a wider population of editors.

Figure 1 illustrates the methods for identifying such personas, with an example of Wikipedia article "Boston". The volume of the number of edits is depicted per quarter and by author. Among top editors, the user Ajd has initially conquered writing the article but sustained his or her activity only to slightly after the mid-age of the article as can also be verified from the rows of user activities. After Ajd, an IP-identified user 67.175.191.237 has had a high peak of edits but only for one quarter – thus, we identify him or her as an independent contributor, a Cowboy. From the other peaks, we can identify sustained contributions from Pentawing and Hertz1888 but Atlant with its one peak of edits we interpret as a Cowboy. Looking at correlations between user editing activities, we can identify users ClueBot and AlexiusHoratus having mostly negative correlations and thus we identify them as Rebels. Visually interpreting the relative edits between users, we can identify a user Loodog as a Follower which has positive correlations with the identified Conquerors with a sustained activity.

## 3. RESULTS

Our research identified four distinct personas, depicted in Figure 2. Our research also provided support for generalization of such personas and differences in editing behaviors of different quality classes in Wikipedia (Table 1). More specifically, in line with the current COINs theories, we found that in FA, there are more Followers than in NA.

We further carried out a statistical testing for the differences in these persona distributions with the Chi-Square test for independence and found the difference to be statistically significant (p<1%), thus rejecting the null hypothesis of equal distribution. We further calculated standardized residuals to measure the degree to which an observed chi-square cell frequency differs from the value expected on the basis of the null hypothesis, finding larger than double standard deviations in the number of Followers and close to standard deviation in the number of Cowboys.

Table 1    Persona distributions in articles

| Article/Author | Conqueror | Follower | Rebel | Cowboy | Total |
|---|---|---|---|---|---|
| Featured | 39 | 23 | 19 | 94 | 175 |
| Non-Assessed | 40 | 6 | 18 | 118 | 182 |
| Total | 79 | 29 | 37 | 212 | 357 |
| Chi-Square | 12,59 | df | 3 | p-value | 0,005613 |
| Percentage distribution: | | | | | |
| Featured | 22,3 % | 13,1 % | 10,9 % | 53,7 % | 100,0 % |
| Non-Assessed | 22,0 % | 3,3 % | 9,9 % | 64,8 % | 100,0 % |
| Standardized residuals: | | | | | |
| Featured | 0,04 | 2,33 | 0,2 | -0,97 | |
| Non-Assessed | -0,04 | -2,28 | -0,2 | 0,95 | |

## 4. DISCUSSION

Our research introduces socially conditioned personas to explain the group dynamics and team creativity, as measured by the article quality in Wikipedia. Our analysis reveals that the leadership role of these personas oscillates during the editing of the article, as previously discovered by Kidane and Gloor (2007) and Kittur (2008). We further discover that there is both order and partial random variation in contribution patterns; this is in line with the virtual team creativity theories (Leenders et al., 2003). The novel contribution of our research is the finding that the editing personalities can be simultaneously individualistic and socially conditioned, manifesting both personal characteristics of the editors and the social reciprocity and dynamics that have an effect on the actualized situated behaviors of these editors. The results can be used in combining the individual and social perspectives in Wikipedia creativity research and social network analysis.

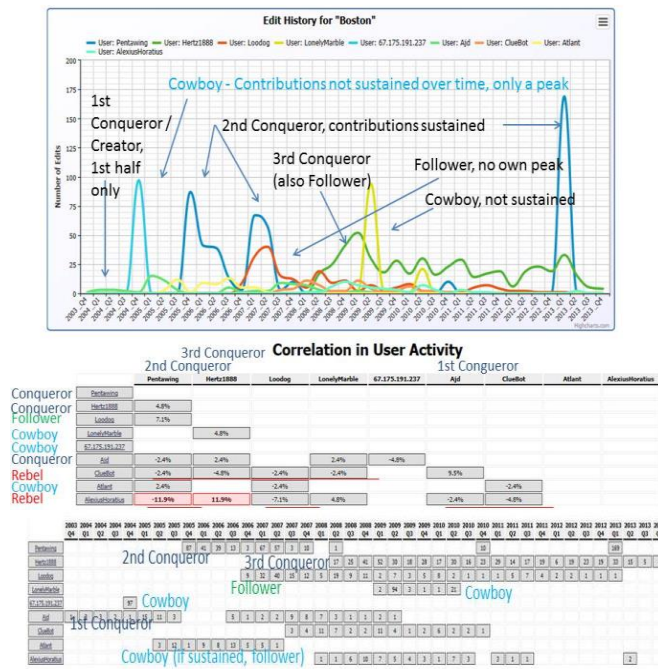

Figure 1    Spotting personas in editing oscillations

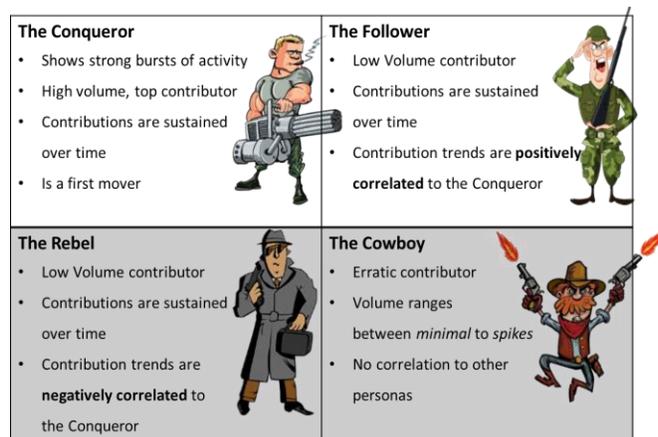

Figure 2    Personas of Wikipedia editors